# High-Power Dual-Channel Field Chamber for High-Frequency Magnetic Neuromodulation

Xiaoyang Tian[*], Hui Wang[*], Boshuo Wang, Jinshui Zhang, Dong Yan, Jeannette Ingabire, Samantha Coffler, Guillaume Duret, Quoc-Khanh Pham, Gang Bao, Jacob T. Robinson, Stefan M. Goetz[#], and Angel V. Peterchev[#]

[*],[#] Equal contributions.

**Abstract**

**Objective.** Several novel methods, including magnetogenetics and magnetoelectric stimulation, use high frequency alternating magnetic fields to precisely manipulate neural activity. To quantify the behavioral effects of such interventions in a freely moving mouse, we developed a dual-channel magnetic chamber, specifically designed for rate-sensitive magnetothermal-genetic stimulation, and adaptable for other uses of alternating magnetic fields.

**Approach.** Through an optimized coil design, the system allows independent control of two spatially orthogonal uniform magnetic fields delivered at different frequencies within a $10 \times 10 \times 6$ cm$^3$ chamber suitable for mouse studies. The two channels have nominal frequencies of 50 and 550 kHz with peak magnetic field strengths of 88 and 12.5 mT, achieved with resonant coil drives having peak voltages of 1.6 and 1.8 kV and currents of 1.0 and 0.26 kA, respectively. Additionally, a liquid-cooling system enables magnetic field generation for second-level durations, and an observation port and camera allow video capture of the animal's behavior within the chamber.

**Main Results.** The system generates high-amplitude magnetic fields across two widely separated frequency channels with negligible interference ($< 1\%$). Relatively uniform magnetic field distribution ($\pm 10\%$ across 94% of the chamber volume) is maintained throughout the chamber, and temperature increase of the inner side of the coil enclosure during the operation is limited to $< 0.35$ °C/s to ensure in vivo safety. Using cobalt-doped and undoped iron oxide nanoparticles, we demonstrate channel-specific heating rates of 3.5 °C/s and 1.5 °C/s, respectively, which validates frequency-selectivity. Both channels can run continuously for four seconds stably.

**Significance.** We present a novel magnetic stimulation platform that combines high-frequency, high-power capability with two independently-controlled channels generating different frequencies, along with a real-time behavioral observation system for freely moving animals. The system supports frequency-multiplexed stimulation strategies for precise modulation of neural activity, making it a versatile tool for advancing magnetogenetics, neural circuit interrogation, and noninvasive stimulation approaches in neuroscience and bioengineering.

## 1. Introduction

Precise control of neural activity is essential for studying brain function and behavior, with applications ranging from fundamental research to potential therapeutic interventions [1-4]. One technique that can enable precise remote control over neural function is magnetothermal genetics [5-8]. This method relies on expressing genetically encoded temperature-sensitive ion channels in target neurons and injecting nanoparticles that heat up in response to magnetic fields at specific frequencies, thereby activating only specific genetically modified neurons in target regions of nanoparticle injection. Unlike traditional techniques such as optogenetics or electrical stimulation, which require physical implants [9-13], magnetothermal genetics is less invasive and allows the controlling energy (magnetic field) to be delivered noninvasively and remotely for both research and therapeutic applications. Other promising approaches that could enable remote neural control via magnetic fields, include nonthermal magnetogenetics [14-16], magnetoelectric nano- and metamaterials that convert the magnetic field energy to local electric potential differences [17, 18], and potential endogenous magnetic responses [19-21].

For effective magnetogenetics stimulation as well as for other biomedical applications mediated by magnetic fields, a system capable of generating a spatially-uniform magnetic field with multiple frequency channels can enable independent simultaneous modulation of different neural targets or subjects [7, 22, 23]. However, existing approaches typically rely on a single-coil setup [7, 22, 24, 25], which generates localized fields and lacks the capability to independently control multiple frequency channels or operate them simultaneously. Moreover, although some previous designs support multi-channel systems across a wide frequency range [22, 26], they have limited control flexibility, as they rely on relays to switch between channels, which precludes simultaneous activation of both channels and limits hardware robustness. The generation of high-power magnetic fields across a wide operational frequency range presents significant engineering challenges for both the electromagnet design and the power electronic system. These challenges are compounded by the need to produce strong and relatively uniform fields within a spatially extended behavioral arena, ensuring sufficient intensity for target activation. High-frequency operation significantly increases the electronic switching losses and the AC losses in the electromagnets. Further, the required high coil current amplitude and resonant voltages increase the heating and stresses within the power switching devices and the coils.

We present a chamber system (Fig. 1) which addresses these challenges with two independently-powered magnetically-orthogonal coils generating strong, sustained, and spatially uniform magnetic fields, allowing a freely moving mouse to be stimulated with two channels simultaneously or sequentially. The system comprises an electromagnetic coil chamber, a coil driver, a DC power supply, a cooling unit, and a behavior observation camera. The electronic coil drive circuitry uses an inverter with an interleaved control strategy to handle high-frequency channels and reduce the switching burden on each transistor, while maintaining the high current-carrying capacity of silicon field-effect transistors necessary for handling the substantial power demands of magnetogenetics applications. The chamber design has inner dimensions of $10 \times 10 \times 6$ cm$^3$ and accommodates a ~ 575 cm$^3$ enclosure for the animal subject, allowing for natural movement and observation of behavioral responses to stimulation. The coil design was optimized to balance peak current and voltage and achieves a sufficiently uniform magnetic field for effective stimulation throughout the enclosure. We implemented a split excitation configuration to reduce the voltage stress on both the coils and compensations capacitors and encased the chamber exterior with high-permeability ferrite tiles to reduce the required energy. Additionally, a water-cooling system was incorporated to support continuous operation, while a compact observation setup enables behavioral monitoring within the chamber.

## 2. Methods

### 2.1. Overview of design

The objective was to generate sinusoidal magnetic fields up to 88 mT at 50 kHz for Channel 1 and 12.5 mT at 550 kHz for Channel 2 to heat up 15 nm cobalt-doped iron oxide nanoparticles (IONP) ($Co_{0.65}Fe_{2.35}O_4$) and 19 nm IONP ($Fe_3O_4$), respectively [22, 23]. The system specifications and coil parameters are listed in **Fehler! Verweisquelle konnte nicht gefunden werden.**.

The proposed magnetic chamber included an acrylic mouse enclosure inside an acrylic casing of the electromagnet coils and coolant and was ensheathed with ferrite tiles of 20 mm total thickness (FPL Series, KEMET Electronics, USA), as presented in Fig. 1. The front of the coil casing formed an opening to allow insertion of the mouse enclosure like a drawer. During operation, the front opening was closed with a ferrite-tiled door to contain the magnetic field and ensure energy efficiency, with two ports installed on the door for air circulation inside the chamber. Due to the high power of the system, a continuous flow of coolant through the coil casing was necessary to maintain safe temperatures for prolonged operation. A water-cooling system was implemented with inlet and outlet tubes connected to an external pump (8020-833-238BX, Pentair, USA) through two unsealed openings at the top of the magnetic chamber that also accommodated the coil terminal wires. Additionally, a hollow conduit (well) was integrated at the center of the chamber's top surface to accommodate the video observation system.

Table 1 System specifications and coil parameters

|  | Units | Channel 1 | Channel 2 |
|---|---|---|---|
| Resonant frequency, nominal | kHz | 50 | 550 |
| Magnetic field flux density (peak) | mT | 88 | 12.5 |
| Winding litz wire strand diameter | mm | 0.1 | 0.05 |
| Winding litz wire number of strands per turn | – | 3160 | 4×500 |
| Winding number of turns | – | 5 | 2 |
| Number of coils | – | 2 | 2 |
| Inductance per coil, first and second halves | μH | 4.4, 4.0 | 1.4, 1.5 |
| Mutual inductance between two halves | μH | 0.7 | 0.4 |
| Calculated compensation capacitance per coil | nF | 2000, 2200 | 47, 44 |
| Implemented compensation capacitance per coil [a] | nF | (6×680)/2=2040, (3×470+3×1000)/2=2205 | (8×6.8+8×4.7)/2=46.0, (6×6.8+10×4.7)/2=43.9 |
| Resonant frequency, measured | kHz | 48.9 | 543 |
| Resonant voltage (peak), measured | kV | 1.5 | 1.8 |
| Excitation current (peak), measured | kA | 1.0 | 0.26 |

[a] For the capacitance implementation, multiplications and additions indicate parallel connection of capacitors, for which their capacitance adds together, whereas the divisions by 2 indicate series connection of the two identical parallel combinations that reduces the total capacitance to half of each combination.

## 2.2. Coil design

The coils for the two channels were arranged to generate orthogonal magnetic fluxes for decoupled operation, minimizing the mutual inductance and preventing interference between channels (Figs. 1 and 2). This design allowed each channel to be controlled independently, without crosstalk. For each channel, the coil was further divided into two halves (coil #1 and #2 for Channel 1 and coil #3 and #4 for Channel 2) to reduce the voltage requirements of the coil connections and resonant drive capacitors. Compared to a single coil for each channel, the inductance of the two independent half windings decreased by a factor of four and the terminal voltage was halved, which reduced insulation requirements and minimized the risk of dielectric breakdown.

To achieve the higher magnetic field strength required for the low-frequency (50 kHz) field in Channel 1, its windings formed the inner layer of the coil assembly, whereas the high-frequency windings constituted the outer layer. Since the front opening of the chamber had to be unobstructed by windings, the chamber there extended beyond the mouse enclosure to accommodate winding segments of Channel 2 (550 kHz) around the opening. Fig. 2A and B respectively show the configuration of the coils for Channel 1 and Channel 2; Fig. 2C visualizes the overall magnetic chamber structure. Finite element analysis results of the magnetic field distribution, simulated in the JMAG software package (JSOL Corporation, Japan), are presented for the two channels in Fig. 2D and E. The coil dimensions were designed to balance resonant voltage and excitation current, ensuring neither was excessively high.

For the lower frequency excitation current in Channel 1, we employed a $3180 \times \varnothing 0.1$ mm litz wire to minimize conduction losses, whereas for the higher frequency current in Channel 2, we used four paralleled $500 \times \varnothing 0.05$ mm litz wires to mitigate AC losses due to the skin and proximity effects. The Channel 1 windings (coils #1 and #2) were wound vertically with five turns each, and the Channel 2 windings (coils #3 and #4) were wound horizontally for two turns.

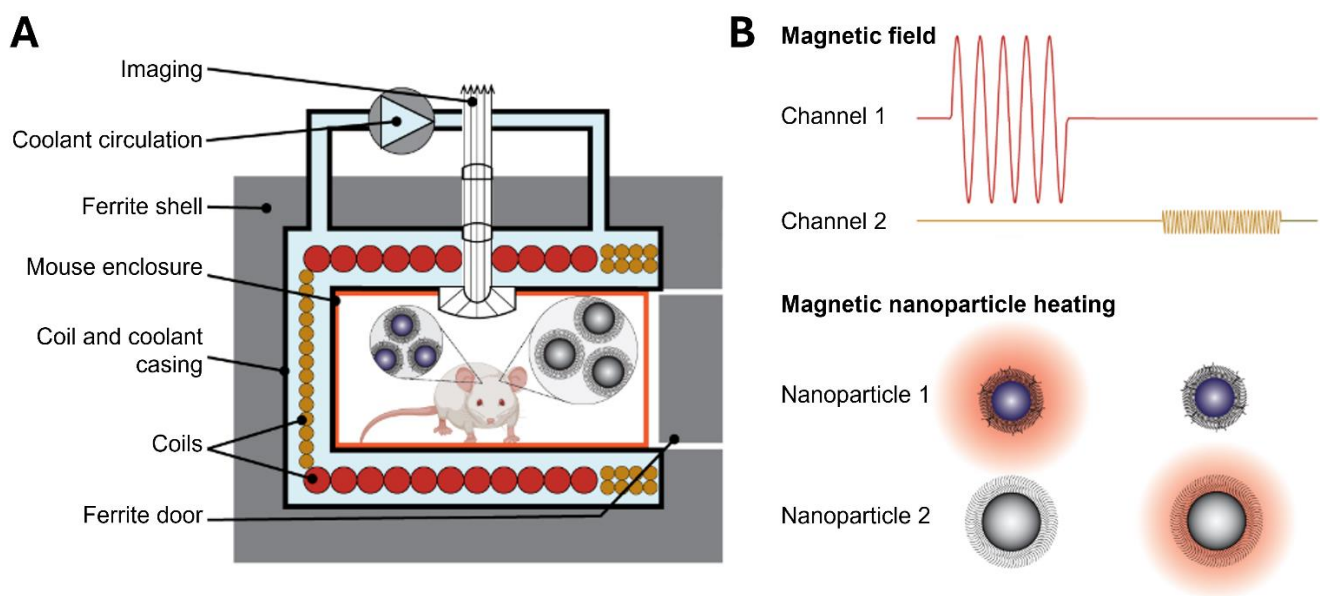

**Fig. 1 Magnetic field chamber diagram and principle of operation. A**. The system enables independent generation and control of two magnetic field channels to selectively heat different magnetic nanoparticles injected within a freely behaving mouse. The chamber integrates a coil assembly with ferrite shielding for high field strength, active liquid cooling for thermal stability, and an optical system for animal behavior observation. The magnetic field of the two channels are horizontally (left to right for Channel 1, red windings) and vertically (Channel 2, yellow windings) orientated, respectively. The mouse enclosure (orange) has a $10 \times 10$ cm$^2$ base size and a 6 cm height and can be inserted via a removable door. **B**. The waveform diagram illustrates how each stimulation channel can produce distinct temporal patterns, allowing control over magnetic nanoparticle activation timing. This platform supports magnetothermal or electromagnetic stimulation studies with spatiotemporal precision.

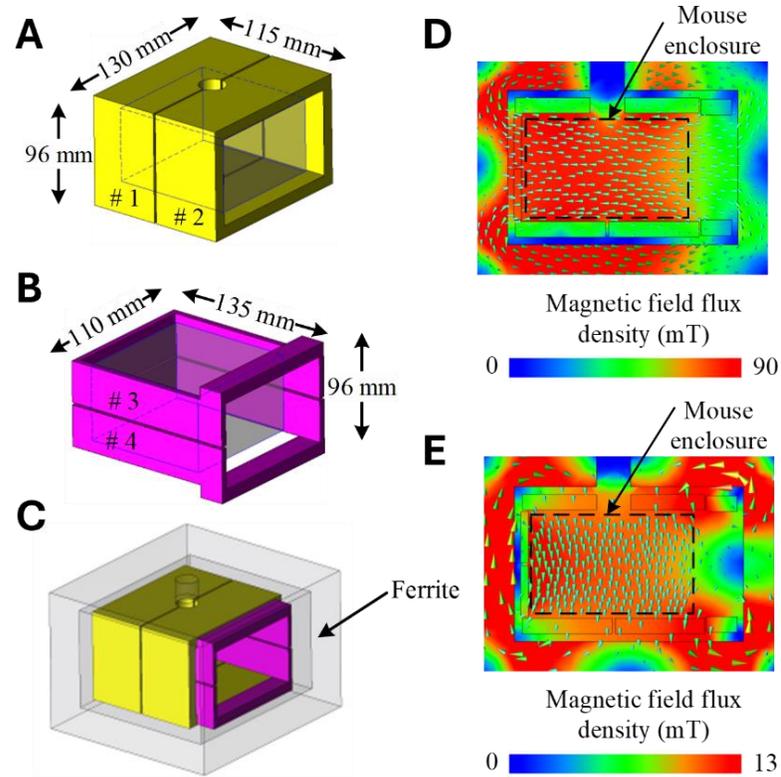

**Fig. 2 Design of magnetic stimulation chamber.** Two independent windings for (**A**) Channel 1 (50 kHz) and (**B**) Channel 2 (550 kHz) are surrounded by a ferrite shell (**C**) and generate spatially orthogonal uniform magnetic fields inside the chamber for Channel 1 (**D**) and 2 (**E**), respectively. The front opening of the chamber via a removable door of the ferrite shell is on the right side of the panels.

## *2.3. Coil driver design*

A diagram of the coil driver is shown in Fig. 3A. One 60 Hz 208 V three-phase outlet sources the main power, and one 115 V single-phase outlet powers auxiliary circuits, including the controller and the transistor gate drivers. To guarantee the magnetic field strength specified in **Fehler! Verweisquelle konnte nicht gefunden werden.** in the working space inside the chamber, the required maximum currents for the two channels were respectively 1.0 kA (Channel 1) and 0.26 kA (Channel 2), and the corresponding input power from AC outlet was 9 kVA and 4.5 kVA, respectively. Below, we describe design details and strategies adopted to optimize the system's performance.

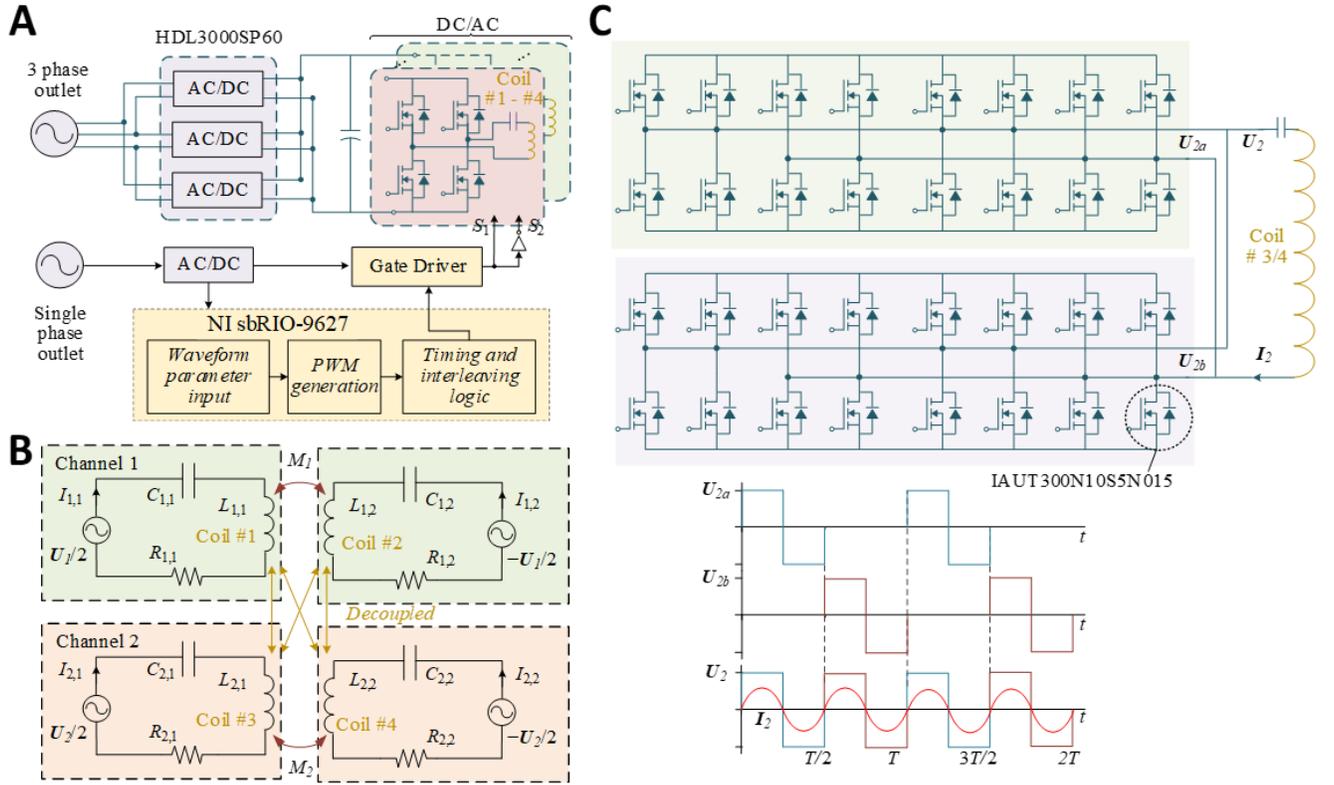

**Fig. 3.** **Power architecture, magnetic channel equivalent circuits, and interleaved drive strategy. A.** System architecture for dual-channel magnetic field generation, requiring both three-phase and single-phase input. The system integrates AC/DC and DC/AC converters, FPGA-based pulse width modulation (PWM) control, and multi-coil magnetic field output, enabling flexible high-power operation with frequency programmability. **B.** Equivalent circuit model of the dual-channel magnetic field generation, with each channel driving a pair of coils. The design includes impedance matching of the coils within channel and decoupling of the coil between channels, to ensure independent field control for each channel. **C.** Interleaved gate-driver topology for Channel 2 at 550 kHz, using multiple SiC full-bridge modules in parallel to reduce switching stress and improve current sharing, where $U_{2a}$ and $U_{2b}$ represent two phase-shifted submodules operating in parallel. The resulting voltage waveform ($U_2$) and inductor current ($I_2$) illustrate reduced ripples and improved power quality at high frequency. The 100% duty cycle is for illustration purposes only; the actual operation was approximately at a maximum of 10% for Channel 1 and 40% for Channel 2.

### 2.3.1. Electronic design

The coil driver was supplied by three single-phase 3 kW AC/DC units (HDL3000PS48, XP Power, Singapore) with inputs connected to the three-phase power outlet in delta configuration and outputs terminated in parallel to a common 48 V DC bus providing 9 kW maximum power. The drivers for Channel 1 and Channel 2 used, respectively, six and four DC/AC inverter boards designed in-house and described previously [22]. To reduce the voltage stress, each channel's coil was split into two halves, with a synchronized control strategy ensuring that the current in each half flows in the same direction and maintains identical magnitude. This approach halved the required output voltage for the DC/AC inverters. For Channel 1, coils #1 and #2 were each driven by three parallel-connected inverter boards connected in series with a compensation capacitor array. For Channel 2, coils #3 and #4 were driven by two parallel-connected inverter boards connected in series with a compensation capacitor array. To minimize stray inductance, wire connections with opposite current flow were paired, which was essential for reducing the energy requirements and electromagnetic interference. Each inverter board contained two H-bridges with a total of sixteen silicon transistors (IAUT300N10S5N015, Infineon Technologies, Germany) that can handle up to 300 A continuous conduction current. The gate drivers for the 160 MOSFETs were powered by an independent 300 W AC/DC power supply (LOP-300-15, Mean Well, Taiwan, R.O.C.).

### 2.3.2. Control strategy

We used LabVIEW (National Instruments, United States) to control in real time the system controller chip with a field-programmable gate array (FPGA) and ARM Cortex A9 microcontroller (Xilinx Zynq 7000, sbRIO-9627, National Instruments, TX, USA). The inverter boards generated biphasic rectangular pulses at each of the channel frequencies, which resulted in sinusoidal coil currents and thus magnetic fields due to filtering by the resonant circuits. The current and field amplitude were specified by the pulse width (duty ratio) and therefore controlled via pulse-width modulation (PMW). For Channel 1, all inverter boards operated in synchrony. To efficiently manage the high-frequency operation and minimize the load on the gate drivers, an interleaved control strategy was implemented for Channel 2 (550 kHz) with two subsets of inverter boards operated alternatingly [22], as illustrated in Fig. 3C. By distributing the switching activity between two inverter sets, the switching frequency for each inverter was halved, which reduced the switching losses and thermal stress on individual inverters and enhanced the overall system efficiency.

### 2.3.3. Compensation circuit

The resonant drive provides a low impedance loop and maximum efficiency [22]. As illustrated in Fig. 3B, although the coils of the two channels were decoupled from each other, the two coil halves within each channel were strongly coupled. Consequently, to ensure the current synchronization and zero-phase operation of the system, the mutual inductance between the two coil halves should also be compensated alongside each coil's self-inductance. Therefore, the compensation capacitor should be calculated according to

$$C_{i,j} = \frac{1}{\omega_i(L_{i,j} + M_i)}, \quad i,j = 1,2 \tag{1}$$

where $C_{i,j}$ and $L_{i,j}$ respectively denote the compensation capacitance and the self-inductance of the first and second ($j$=1,2) coil halves of channel $i$, $M_i$ represents the mutual inductances between the two coil halves, and $\omega_i$ is the angular frequency of channel $i$. The self- and mutual inductances were estimated through JMAG simulation and calculated by the terminal voltage-current ratio (**Fehler! Verweisquelle konnte nicht gefunden werden.**) as

$$\begin{cases} L_{i,j} = \dfrac{U_{i,j}}{\omega_i \cdot I_{i,j}} \\ M_i = \dfrac{U_{i,(3-j)}}{\omega_i \cdot I_{i,j}} \end{cases}, \quad i,j = 1,2 \tag{2}$$

where $U_{i,j}$ denotes the terminal voltage of the $j$-th coil half of channel $i$. For example, when the first coil half of Channel 1 was charged with current $I_{1,1}$, the voltage of the first coil half $U_{1,1}$ was proportional to the coil inductance $L_{1,1}$, while the voltage of the second coil half $U_{1,2}$ was proportional to the mutual inductance $M_1$.

The compensation capacitances were calculated and implemented by combining $2 \times 6$ and $2 \times 16$ capacitors (MKP, WIMA, Germany) in a configuration of two parallel-connected combinations in series for Channel 1 and Channel 2, respectively, to distribute high voltage and high current load. Specifically, $C_{1,1}$ consisted of twelve 680 nF capacitors, $C_{1,2}$ consisted of six 470 nF and six 1000 nF capacitors, $C_{2,1}$ consisted of sixteen 6.8 nF and sixteen 4.7 nF capacitors, and $C_{2,2}$ consisted of twelve 6.8 nF and twenty 4.7 nF capacitors (see **Fehler! Verweisquelle konnte nicht gefunden werden.**).

### *2.4. Optical observation system*

Since the magnetic chamber was fully enclosed with ferrite tiles to minimize energy use and emission of electromagnetic interference, a custom observation system was required to monitor mouse behavior within this closed environment. Therefore, the top of the chamber incorporated an 18-mm-diameter well nestled between coils #1 and #2 to accommodate lighting and real-time observation inside the mouse enclosure. To allow a wide-angle view of the mouse chamber with its significant side width of 100 mm and limited height of 60 mm, and considering the fact that the subject (a young-adult mouse) is typically ~ 30 mm high and tends to stand up, a fisheye lens (195° Super Fisheye Lens, HD5V2, APEXEL, China) was installed in the chamber, occupying ~ 25 cm$^3$, and coupled to the well (Fig. 4). The lens allowed a wide field of view ($100 \times 100$ mm$^2$) at a limited working distance ($< 10$ mm) due to the aforementioned factors. The lens was custom made with its enclosure material changed from metal to plastic, for compatibility with the alternating magnetic field. Two additional field lenses (SM05 achromatic doublet and triplet, AC127-019-A, TRS127-020-A, Thorlabs, USA) and one relay lens (TRS254-040-A, Thorlabs) were placed in a 3D-printed plastic lens tube (Black V4 resin, Form 3+, Formlabs, USA) to collect and transmit the light from the fisheye lens to a camera (BFLY-PGE-14S2C, Teledyne FLIR, USA) outside the chamber, at a safe distance from the magnetic fields. The compact lens assembly, mounted using low-profile threading supports (tube-to-camera: external CS-mount to external Thorlabs SM1 adapter; tube-to-fisheye: internal & external Thorlabs SM05 threading), ensured minimal intrusion and mechanical stability. The combination of two 0.5-inch (12.7 mm) field lenses and a 1-inch (25.4 mm) relay lens effectively collected and transmitted light ray output from the fisheye lens, enabling a clear top-down view. The design supported high-speed video acquisition and precise behavioral analysis, even in the presence of strong alternating magnetic fields. In addition, four small flat LED lights (LX18-P140-3, Lumileds, Netherlands) were installed on the ceiling of the chamber to provide uniform illumination across the mouse enclosure. The wires powering the LEDs were twisted to minimize interactions with the magnetic field and localized heating and were passed through the well along the fisheye lens shaft.

### *2.5. Electrical and magnetic field measurements*

We performed all electrical measurements with a 500 MHz high-bandwidth oscilloscope (Tektronix MDO3054, Beaverton, OR, USA). We used high-voltage differential probes (Tektronix THDP0100) and Rogowski current transducers (CWTUM/60, Power Electronic Measurement Ltd., Nottingham, UK) to measure the coil and capacitor voltages and currents, respectively. We created a magnetic field sensor by winding ten turns of 30-gauge magnetic wire around a 1 cm diameter cylindrical former 3D-printed from polylactic acid [22]. The sensor measured the voltage induced by the changing magnetic flux of the alternating magnetic field, from which the field strength is derived.

We confirmed each channel's resonance frequency by varying the electronic circuit's output frequency and maximizing the measured coil current and voltage, or the induced voltage from the magnetic field probe. Once confirmed, the resonance frequency remained fixed throughout operation. We used a handheld infrared thermometer (Fluke Corporation, Everett, WA, USA) to detect the temperature of the coils, cable, and driver boards.

### *2.6. Observation of mouse inside coil chamber*

All animal handling and experiments were performed in accordance with Rice University IACUC guidelines (IACUC-23-038-RU). The ten-week-old mouse (c57BL/6, Jackson Laboratory, USA) was transferred from the cage to an acrylic chamber with outer dimensions of $10 \times 10 \times 6$ cm$^3$ and a perforated top lid to allow air circulation. The chamber was then placed inside the magnetic chamber and images were captured in FlyCapture2 software (Teledyne FLIR) using the optical platform presented in Fig. 4.

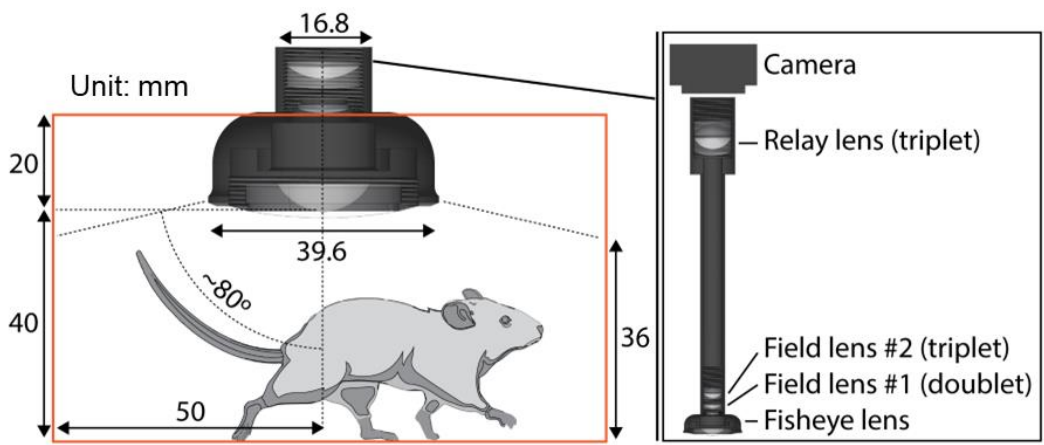

**Fig. 4 Optical mouse behavior observation system.** The system uses a fisheye lens combined with a relay optical path to capture the interior of the magnetic chamber, enabling real-time tracking of a freely moving mouse during dual-channel magnetic stimulation. The volume in which the mouse is fully visible from the lens is about 370 cm$^3$.

*2.7. Nanoparticle synthesis and measurements*

We used 15 nm cobalt-doped IONP for low-frequency high-amplitude magnetic field (Channel 1) and 19 nm IONP in a high-frequency lower-amplitude field (Channel 2), with their syntheses and characterization following previous methods [22, 23]. In summary, magnetite nanocrystals were synthesized through thermal decomposition and grown to designated sizes by controllable seed-mediated growth. Following characterization with a transmission electron microscope, a dual-solvent exchange method was employed to coat nanoparticles with polyethylene glycol (PEG2K). The nanoparticle samples were placed inside the coil chamber and the heating was measured using the handheld infrared thermometer and fiber optic thermal probes (Luxtron 812 and STF-2M probe, Lumasense, Santa Clara, CA, USA). The specific absorption rate (SAR) was calculated as

$$\text{SAR} = \frac{C}{\rho} \cdot \frac{\Delta T}{\Delta t},$$

where $C$ is the specific heat capacity of the media (4180 J·kg$^{-1}$·K$^{-1}$), $\Delta T$ is the temperature change during stimulation averaged over three stimulations, $\Delta t$ is the stimulation time, and $\rho$ is sample density measured by total metal concentration. The undoped and cobalt-doped iron oxide nanoparticles were recorded at concentrations of 21.46 mg$_{metal}$/mL and 20.68 mg$_{metal}$/mL, respectively. Thermal images were recorded with an infrared camera (FLIR A700, FLIR, Wilsonville, OR, USA), which was unaffected by magnetic field.

## 3. Results

*3.1. Implementation of coil chamber, electrical performance, and magnetic field*

The implementation of the system is shown in Fig. 5A. The DC power supply unit contained three AC-DC converters that received three-phase 208 V AC input and provided the main power to the coil driver unit at 48 V DC. The coil driver unit (Fig. 5B) contained the FPGA controller, coil driver boards, compensating capacitors, and auxiliary components (additional DC power supplies, cooling fans). It received control from LabVIEW running on a computer and operated the two sets of coils in resonance. The coil enclosure (Fig. 5C and D) contained the two coils and ferrite shielding to generate high frequency alternating magnetic fields inside the magnetic chamber. The water-cooling pump circulated water to cool the coils.

We verified that the size of the mouse chamber could accommodate a mouse and allow it to move freely (Fig. 5E). The observation system worked as expected: through the fisheye lens, we can observe various behaviors of the mouse, including resting, circling, and raising its head (Fig. 5F).

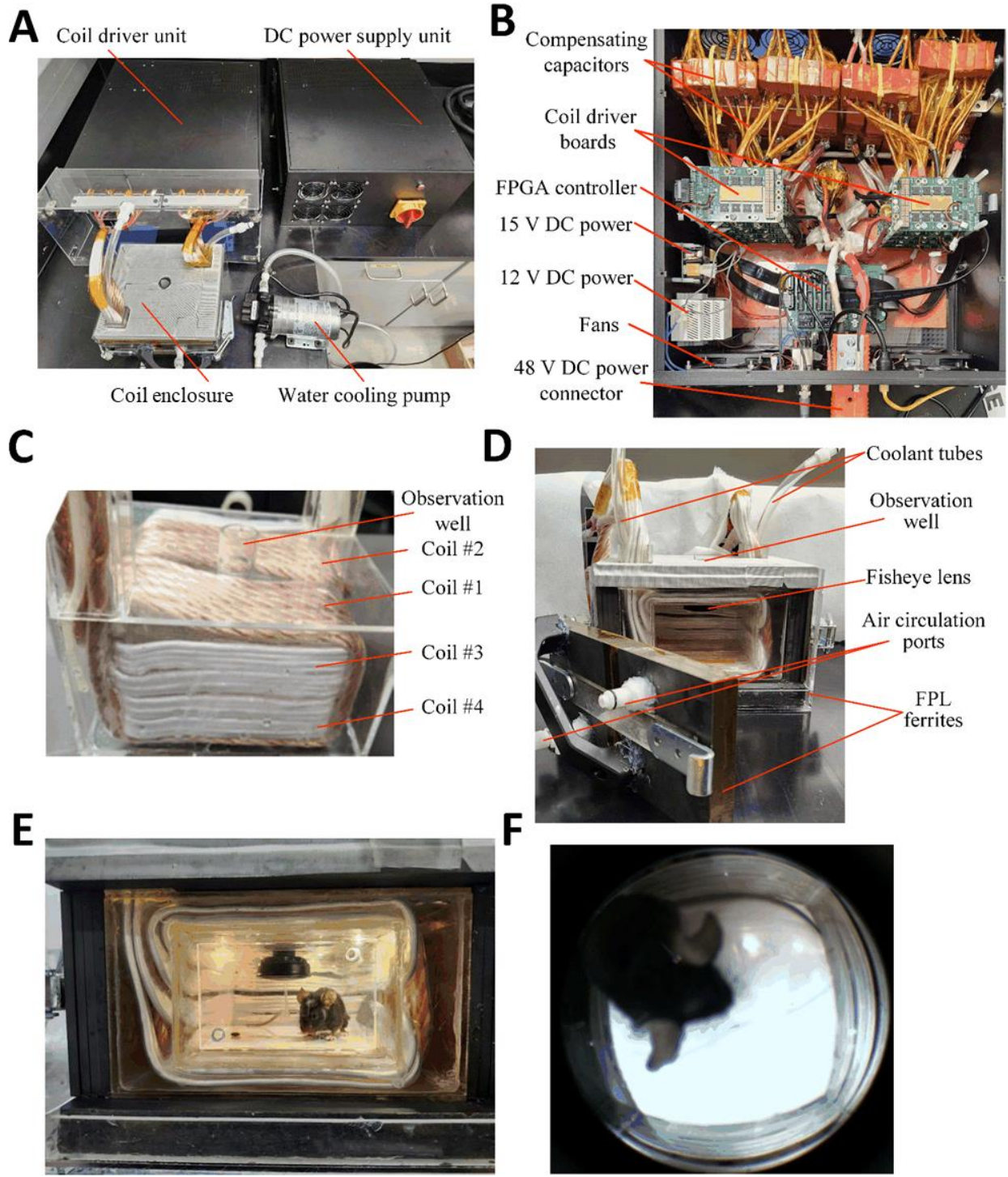

**Fig. 5 Experimental prototype. A.** Whole system. **B.** Components inside coil driver unit. **C.** Back view of coil enclosure without coolant and surrounding ferrite. **D.** Front view of the coil enclosure with ferrite shielding, coolant tubes, fisheye lens installed. The removable front door has a handle and ports for connecting air circulation. **E.** A mouse inside the chamber, illuminated by LEDs on the chamber ceiling. **F.** Raw image of the mouse inside the chamber captured via the fisheye lens.

For Channel 1, the system was resonant at 48.9 kHz, and the mouse chamber's magnetic field density reached 88 to 110 mT when coil #1 and coil #2 were driven with 1 kA peak (Fig.

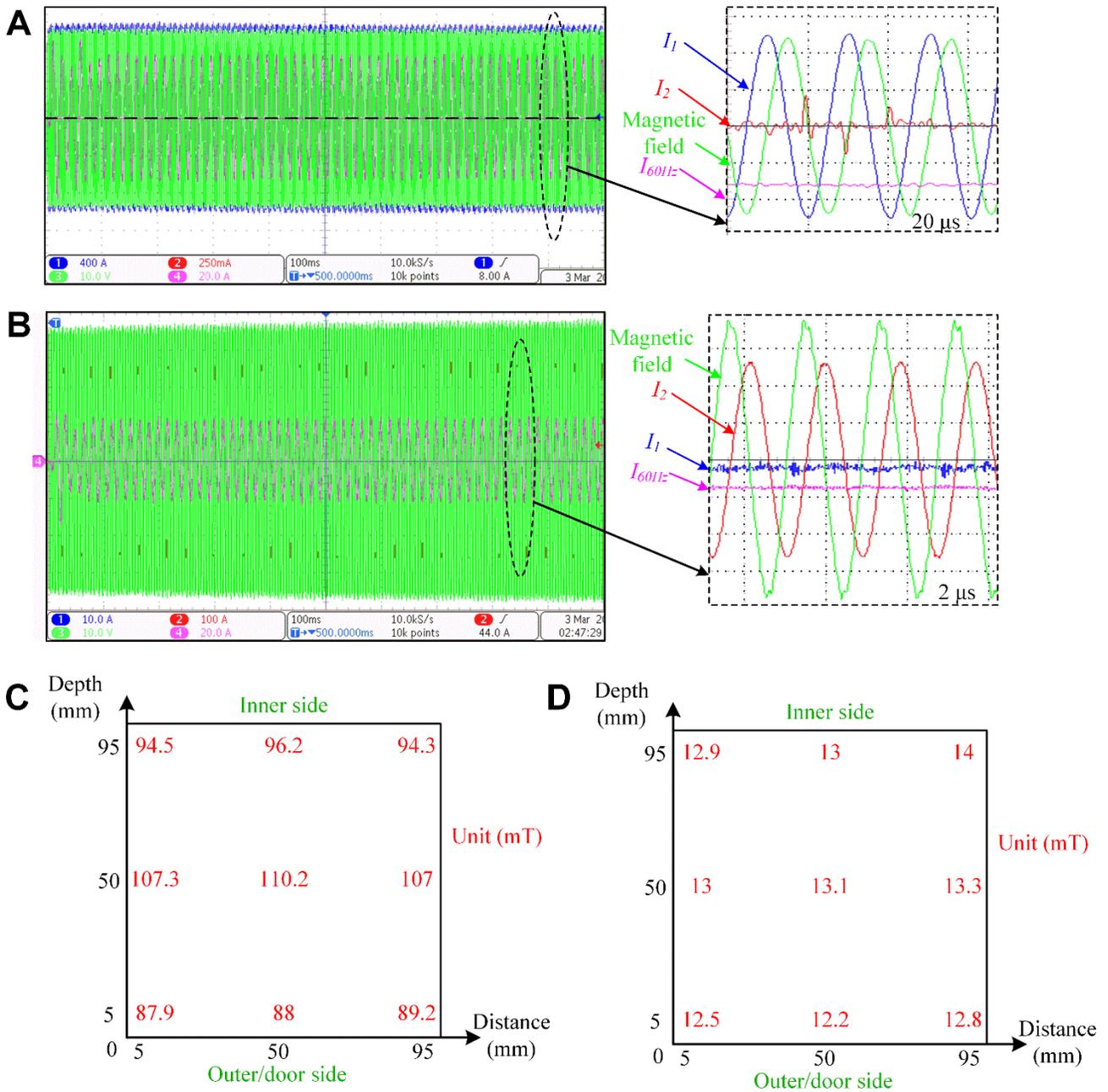

**Fig. 6**A). Although the magnetic field near the door was slightly weaker compared to that deeper inside the chamber (Fig.

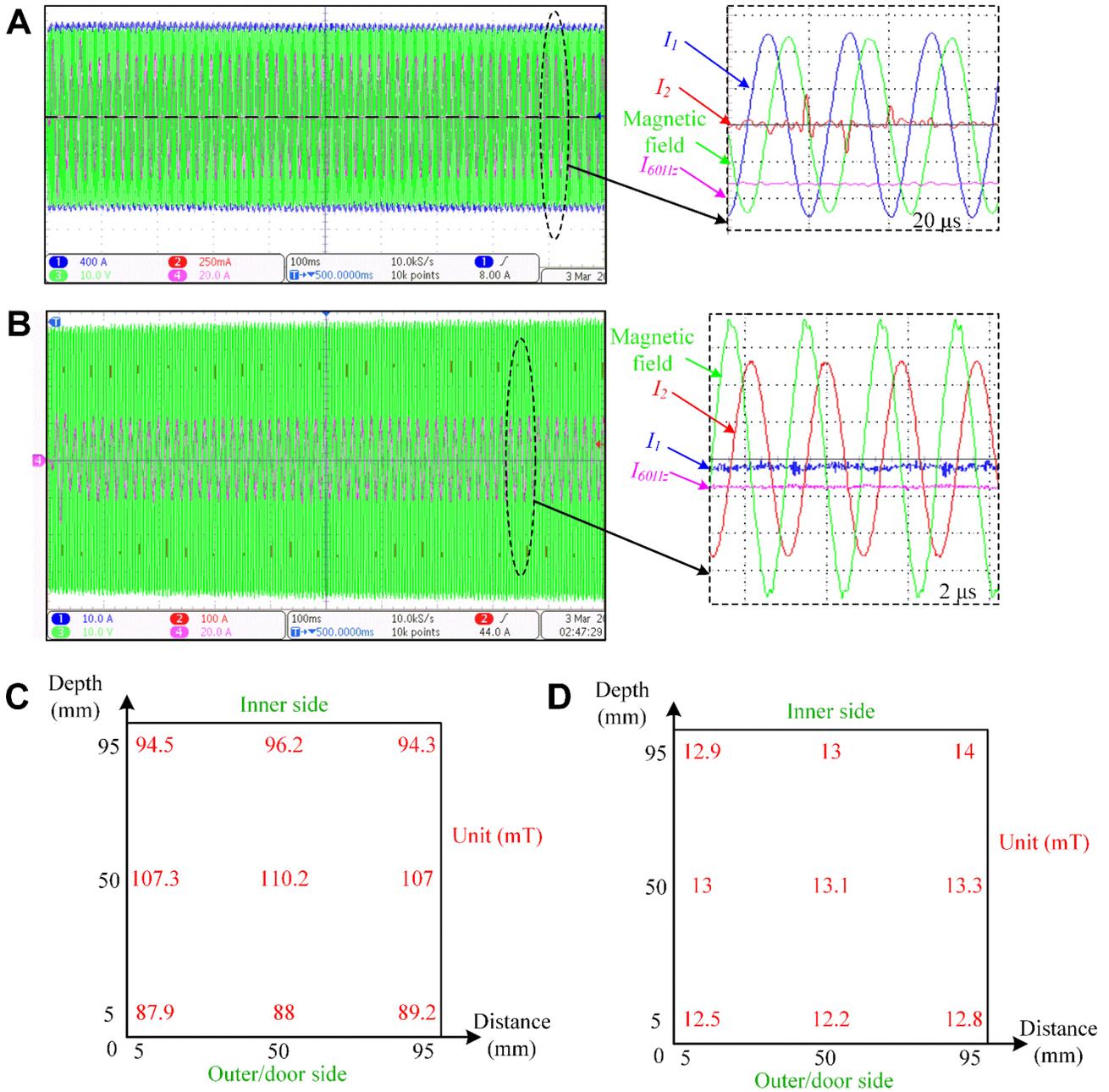

**Fig. 6**C), it still met the specification of 88 mT. For Channel 2, the system was resonant at 543 kHz. As shown in Fig.

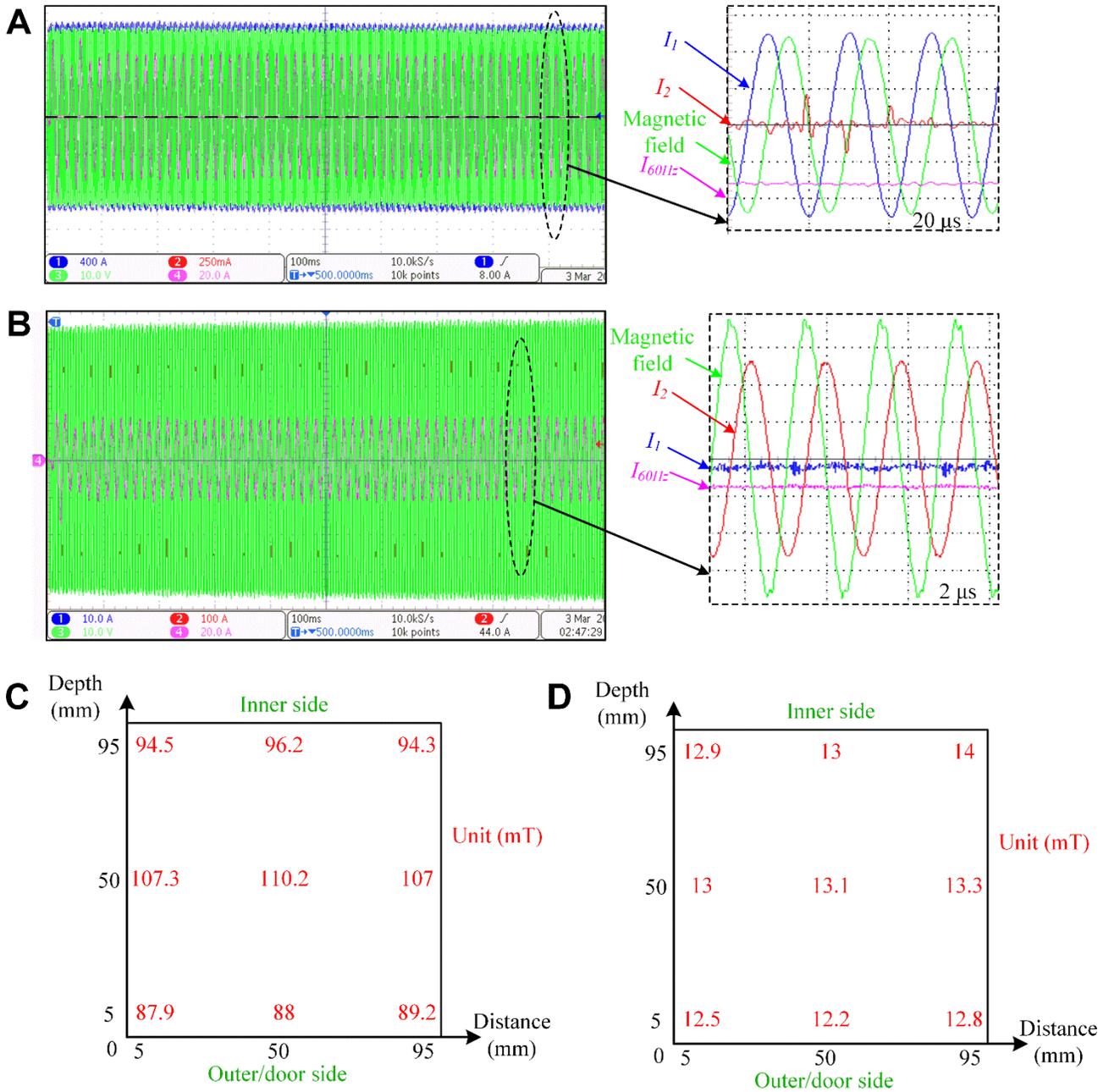

**Fig. 6**B, when coil #3 and coil #4 were driven with 260 A peak, the magnetic field inside the mouse chamber was quite uniform, 13 ± 1 mT (Fig.

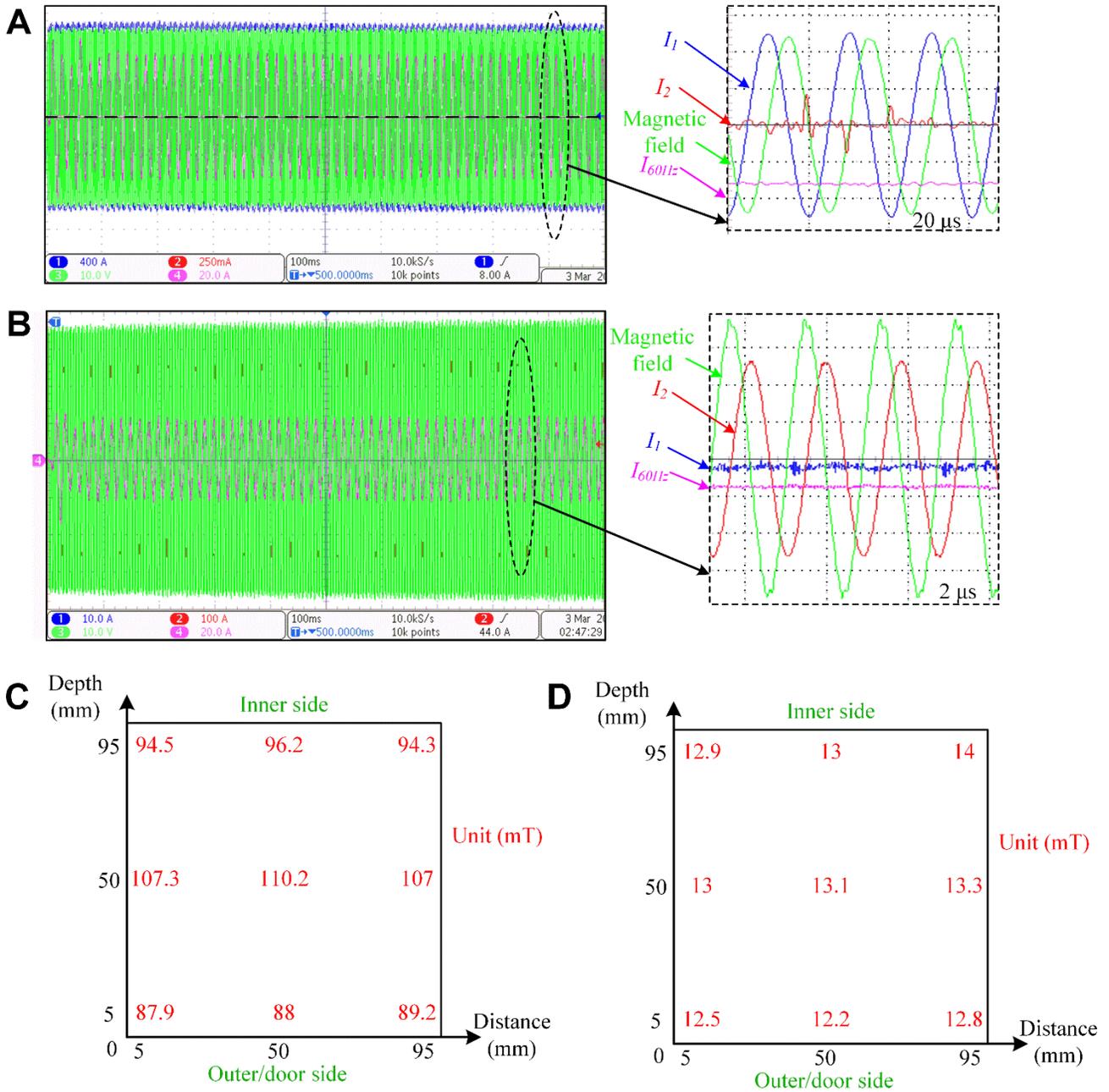

**Fig. 6**D). The typical operation waveforms (Fig.

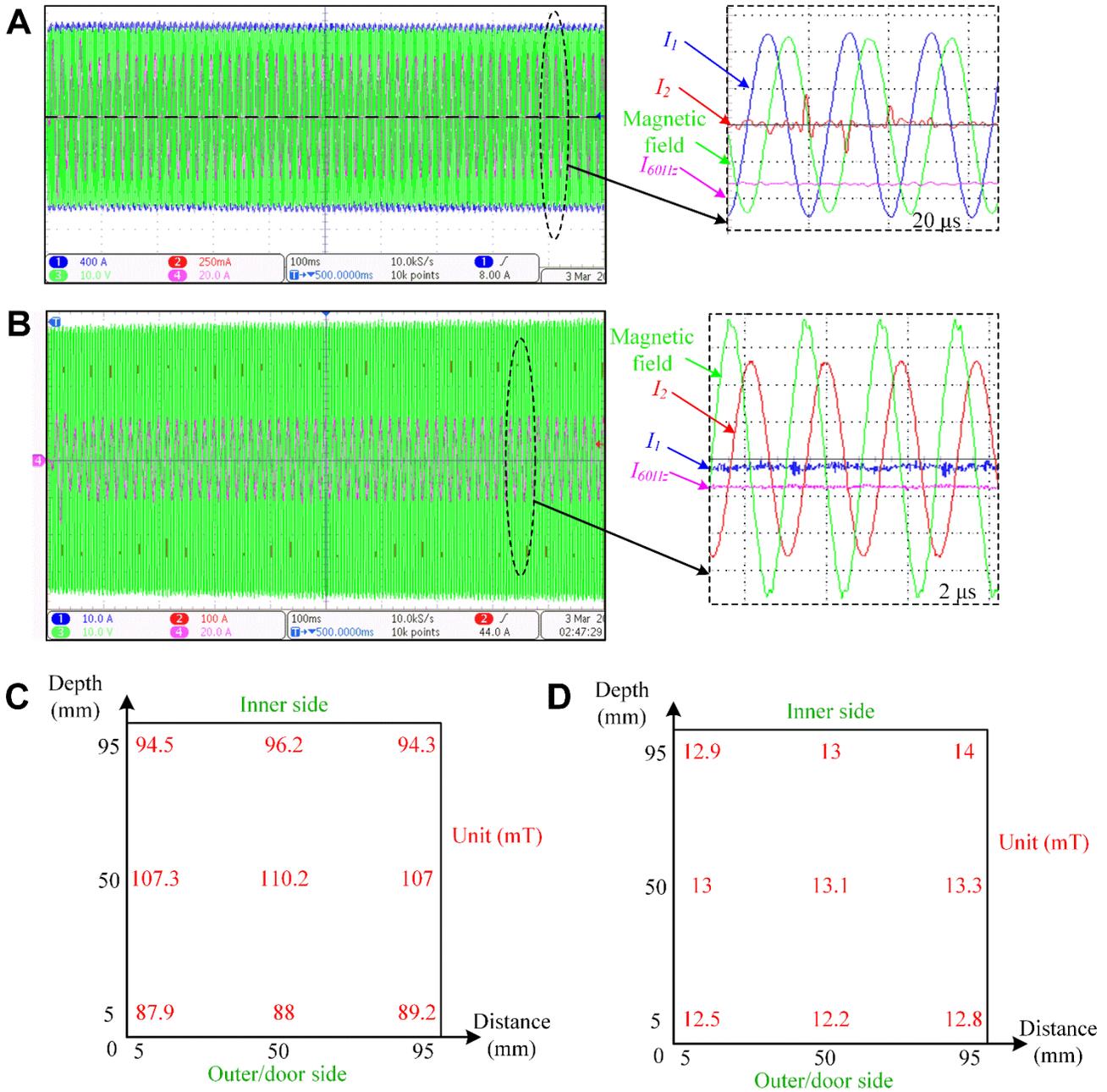

**Fig. 6**A and B) also indicate that when one channel was running, the current waveform of the other channel was almost zero, indicating minimal coupling and interference between them (< 1% crosstalk), as intended. In addition, the three-phase input currents for Channel 1 and Channel 2 were respectively 25 A and 14 A (root-mean-square value). The input currents confirmed that the 9-kW power unit met the power requirement of either channel.

### 3.2. Thermal characteristics of mouse chamber

Another crucial aspect for animal experiments was to limit the temperature increase inside the chamber. Since Channel 1 carried about 1 kA current, the thermal effect was significant. As shown in Fig. 7A, the litz wire's temperature increased from 29.3 °C to 38.9 °C after 2 s of operation. Nonetheless, the water-cooling system can control the temperature of the chamber within a reasonable range. At the water outlet tube, the temperature increase was only 0.1°C, while the internal surface temperature of the coil enclosure changed by only 0.7 °C (Fig. 7B). For Channel 2, whose current was only 260 A, the temperature change was negligible (< 0.2 °C/s).

### *3.3. Nanoparticle heating*

Finally, we tested the system's heating performance with 15 nm cobalt-doped IONP and 19 nm undoped IONP (Fig. 8). When we applied the Channel 1 magnetic field, the cobalt-doped IONP heated at a rate of 3.5 °C/s, whereas the undoped IONP only heated at a rate of 0.2 °C/s. In contrast, when we applied the Channel 2 magnetic field, the cobalt-doped IONP heated at a rate of only 0.1 °C/s, whereas the undoped IONP heated at a rate of 1.5 °C/s. The selective heating of these two different types of IONPs highlights the potential to perform multi-channel magnetothermal genetics research in behaving mice.

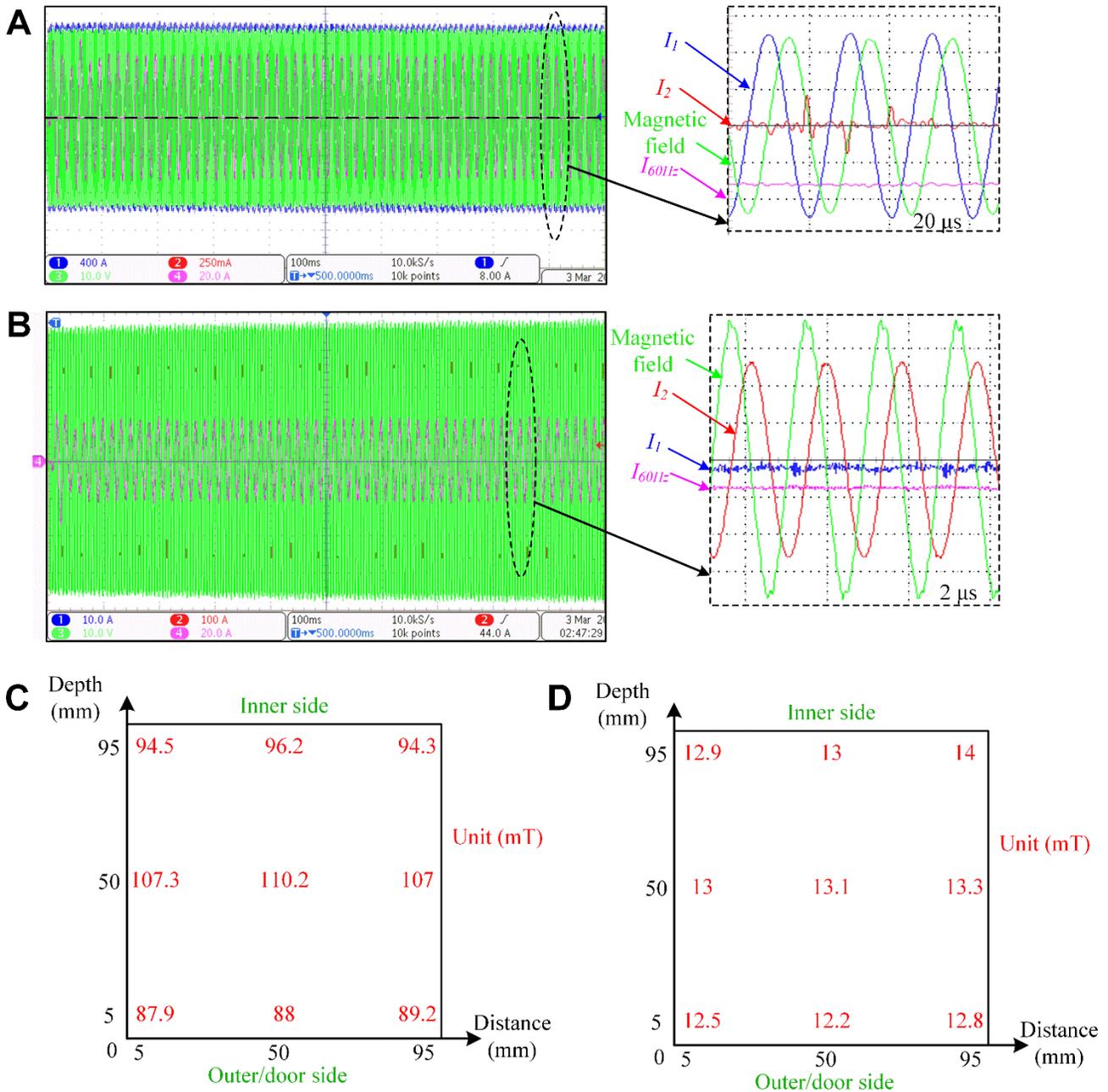

**Fig. 6. Electromagnetic measurements.** Current and magnetic waveforms of the system for **(A)** Channel 1 and **(B)** Channel 2. $I_1$ and $I_2$ are the currents of the Channel 1 and Channel 2 windings, and $I_{60\text{Hz}}$ is the input current of one of the phases from the three-phase 60 Hz, 208 V power supplies. Zoomed-in panels share the same y-axis scale as indicated by the voltage or current per division in the corresponding main panels. Measured magnetic field distribution inside the chamber for **(C)** Channel 1 and **(D)** Channel 2.

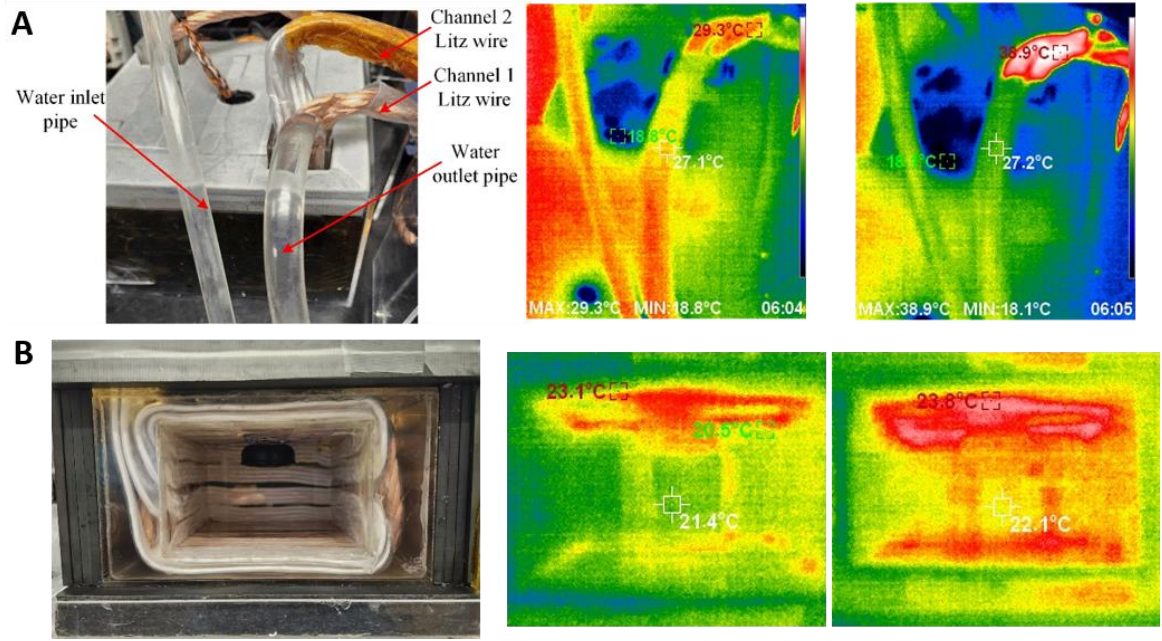

**Fig. 7**. **Temperature changes of the magnetic chamber for Channel 1 operation. A.** Top view of the chamber. **B.** Front view of chamber inside. Middle and right panels respectively show temperatures before operation and after 2 s of operation of Channel 1 at 1 kA.

## 4. Discussion

This paper presented a dual-channel magnetic field generator designed for magnetothermal genetics experiments in mice. The system can produce a strong and relatively uniform magnetic field of 88 mT at 50 kHz and 12.5 mT at 550 kHz in a $10 \times 10 \times 6$ cm$^3$ volume behavioral arena. Compared to prior e-coil or time-multiplexed systems, this work presented an architecture allowing independent simultaneous control over the two channels with negligible interference between channels exceed the capabilities of previously studied dual-frequency systems that often suffer from complicated compensation networks (e.g., CLCL) [27] or additional switches [28], which will add impedance and may break down under the high voltages (1.8 kV) in this system.

The electromagnet coil driver output power reached about 7.7 kW and the measured input power is only 9 kVA (8.8 kW), compatible with standard three-phase power outlets. For the presented AC/DC power supply implementation (9 kW) the channels can be operated sequentially at the maximum magnetic field strengths or simultaneously at lower magnetic field strengths. The system can operate both channels at the maximum magnetic field strength if a more powerful AC/DC power supply or additional parallel supplies are connected to feed the 48 V DC bus.

Thermal management is a critical concern in high-power resonant magnetic systems. In this implementation, despite instantaneous power transfer in the kilowatt range, the internal wall of the chamber temperature rise was kept below 0.35 °C/s during the magnetic field delivery. This limited temperature rise was achieved via an active water-cooling loop surrounding the ferrite-encased chamber. Importantly, the low thermal gradient between the coil terminals and the internal enclosure (typically < 1 °C during 2 s stimulations) demonstrated the mechanical and thermal robustness of the design, supporting safe and repeatable operation across trials. In future experiments with mice, we expect that a 4 °C increase in nanoparticle temperature is sufficient to drive neural activity via genetically expressed heat-sensitive ion channels [23]. Thus, based on the measured rate of temperature change or comparing SAR of the nanoparticles with those of previous studies [23], this system is expected to only need to run Channel 1 for 1.5 seconds or Channel 2 for 3 seconds to deliver a single stimulus.

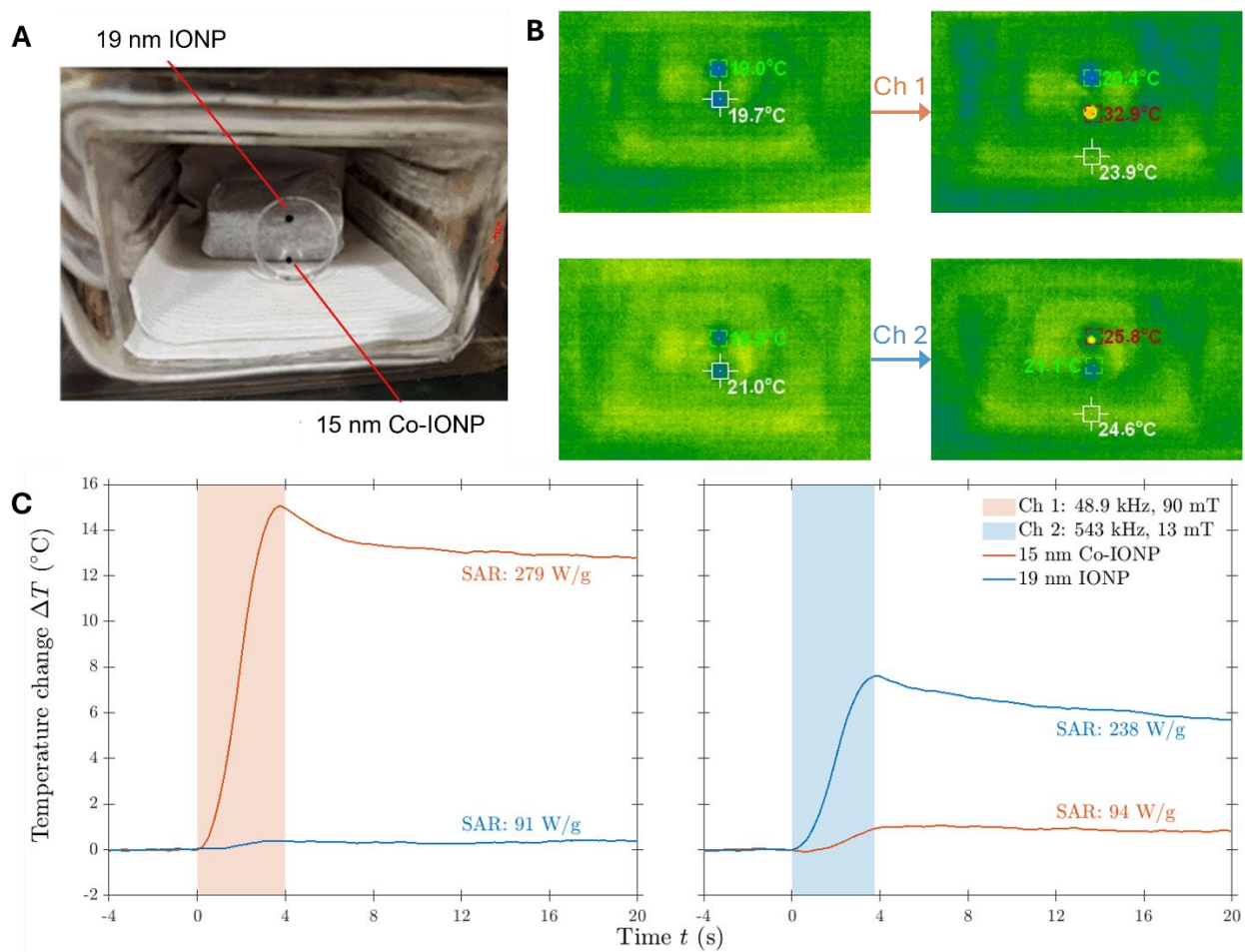

**Fig. 8**. **Nanoparticle heating. A.** Measurement of 10 μL of 19 mm IONP (top) and 10 μL of 15 mm Co-IONP (bottom) in a Petri dish using handheld infrared thermometer and with the chamber door open. **B.** Infrared thermometer reading of nanoparticle temperatures before and after 4 s of operation of Channel 1 (top) and Channel 2 (bottom). **C.** Measurement using the insertion temperature sensor, performed under normal operation condition with the chamber door closed. Responses of the 15 nm cobalt-doped iron oxide nanoparticles (red lines) and the 19 nm undoped iron oxide nanoparticles (blue lines) when exposed respectively to a magnetic field of Channel 1 (right, 48.9 kHz; 90 mT) and Channel 2 (left, 543 kHz; 13 mT).

These features made the system suitable and adaptable for various experimental setups involving magnetic nanoparticles or other magnetically sensitive constructs for both basic and preclinical research. The size of the chamber can accommodate one young mouse. In principle, the chamber can be expanded to accommodate two mice, although such expansion will require further optimization of the coil design to produce a sufficiently uniform field over a wider area, and the power requirements will increase proportionally to the chamber volume, which requires upgrading and further optimizing the driver system.

A further innovation was the integration of a real-time optical observation system. By embedding a fisheye lens observation system and LED lighting inside the chamber, the chamber maintained almost complete ferrite shielding (except for the three openings on the top and air circulation ports in door) while still enabling behavioral video capture. Checkerboard calibration can be performed for the imaging system to remove the fisheye-lens distortion, which is a consequence of achieving large angle-of-view at a close distance. The distortion correction can make the video more interpretable to an observer and better suited for behavior tracking software.

## 5. Conclusion

We developed a dual-channel resonant magnetic field system for magnetothermal genetic stimulation. By properly arranging two orthogonal coils and ferrites, the system can independently generate two powerful magnetic fields within a $10 \times 10 \times 6$ cm³ chamber, and peak coil excitation currents are respectively limited to 1.0 kA and 0.26 kA. By splitting each channel coil into two short coils, the maximum coil terminal voltages are only 1.5 kV (Channel 1) and 1.8 kV (Channel 2). Thermal rise of the chamber was maintained below 0.35 °C/s through an active water-cooling system, ensuring chamber safety and thermal stability during repeated high-power stimulation cycles. Finite- element modeling confirmed a field uniformity deviation within ±10% across the most stimulation volume. The experiment results proves that we can selectively and effectively heat two kinds of nanoparticles, and we can observe mice behavioral through the specially designed and integrated fisheye-lens.

This system provides a scalable and precise tool for multichannel magnetic neuromodulation and represents the first implementation supporting sub-second, independently addressable magnetic stimulation at distinct frequencies in unrestrained small animals. The platform is readily extendable for applications involving magnetoelectric coupling, wireless thermal control, or real-time closed-loop behavioral paradigms. Future work will focus on extending the system to additional frequency channels, multi-animal experiments, and integration with electrophysiological and imaging readouts.

**Acknowledgments**

Research reported in this manuscript was supported by the National Institute of Neurological Disorders and Stroke of the National Institutes of Health under Award Number RF1 NS126063. The content is solely the responsibility of the authors and does not necessarily represent the official views of the National Institutes of Health.

The authors thank Maya E. Clinton, who assisted in building the system, David L. K. Murphy, who provided valuable advice on safety issues and coil manufacturing, and Duke University Co-Lab Studio, which provided water cutting and 3D printing equipment.

**Author contribution**

X.T. and H.W. designed and implemented the coil chamber and driver system, B.W. contributed to the coil design, H.W. and J.Z. developed the control software, H.W. performed measurements on the coil and driver system, D.Y. designed and implemented the optical observation system, S.C., Q.K.P., and B.G. synthesized the nanoparticles, J.I., S.C., and G.D. performed the animal experiment and nanoparticle measurements, J.T.R secured funding and resources for the study and supervised the Rice University team. S.M.G. and A.V.P. conceived, supervised, and secured funding and resources for the study at Duke University. X.T. and H.W. took the lead in writing the manuscript, with inputs from B.W., A.V.P., J.I, S.C., and D.Y. All authors reviewed, commented on, and approved the final version of the manuscript.

**Conflict of interest**

J.T.R. receives monetary and equity compensation from Motif Neurotech. S.M.G. is inventor on patents and patent applications on transcranial magnetic stimulation technology and has received consulting fees from Ampa, royalties from Rogue Research, TU Muenchen, and Porsche, and research support and patent fee reimbursement from Magstim. A.V.P. is an inventor on patents and patent applications on transcranial magnetic stimulation technology and has received patent royalties and consulting fees from Rogue Research; equity options, scientific advisory board membership, and consulting fees from Ampa Health; equity options and consulting fees from Magnetic Tides; consulting fees from Soterix Medical; equipment loans from MagVenture; and research funding from Motif. The other authors declare no relevant conflict of interests.